# Prediction in Cyber Security

*Complications and Consolations*

Antonio Roque
MIT Lincoln Laboratory
Lexington, MA, USA

*Abstract*—Uncertainty, error, and similar complications add to the many challenges of cyber security. Various disciplines have developed methods for managing these complications, but applying these methods involves disambiguating overlapping terminology and determining a method's proper usage in the context of cyber security, which has unique properties. This process is here guided by the need for *prediction*, which is required for cyber security to become more like traditional sciences. A motivating malware analysis example is defined. A five-element ontology of complications for prediction is presented, and concepts from numerous disciplines are mapped to it in terms of the motivating example.

## I. MOTIVATION

Imagine we are asked to analyze a recently-isolated malware sample. A typical approach is to perform a static analysis of the binary involving disassembly and decompilation, or to run a dynamic analysis in which the malware is executed in an isolated environment [55]. Imagine we are then asked: "how bad is it for us? How will the malware affect our organization? What about organizations similar to ours? What about systems in general around the world?" We are being asked to estimate, to generalize, to predict.

What can we claim? How do we express and justify these claims? What are the limits to the claims we can make? Being able to answer these questions is one of the practical requirements for the development of a *science of cyber security* [23][36][56]. Hatleback [21, pp5-6] calls "lack of predictive ability" one of the characteristics keeping cyber security from becoming a true science, citing Schneider [53], Maxion [34], and Kott [30]. But in previous work we have described several unique features of cyber security and how those features make cyber security different from traditional sciences [47], which complicates the issue. Formalizing predictive mechanisms in cyber security remains an open research task.

To understand the nature of prediction in computer security, this paper systematically examines the Complications that arise when performing such a malware analysis and the Consolations that various disciplines have developed to manage these difficulties. This paper was developed in the following way. First, a variety of scientific, engineering, and operational disciplines were surveyed to identify the ways that they express and manage error, uncertainty, and similar complications to prediction. Second, these complications were considered in terms of their manifestation in cyber security, guided by the malware analysis scenario described above.

Finally, the complications were organized in the ontology shown in Table I. The next section, which makes up the greater part of this paper, discusses this ontology.

## II. ONTOLOGY

Intuitively, there are a number of possible problems when performing malware analysis. To begin with, our disassembler, decompiler, or execution environment might have undocumented bugs. So one aspect of uncertainty is related to the *Tool* being used.

And even if our tools are bug-free, we might use them the wrong way. We might use numbers and statistics to make incorrect inferences, which would be a problem of *Method*. Or we might mis-read or mis-type a number, or be subtly biased in our interpretation of the results, in which case the problem would be in the *Analyst*, namely ourselves.

Even without that, there might be a problem caused by the *Adversary*, who might react in an unexpected way. Finally, there might be unknowable issues in terms of the *Context*.

These are all *aspects* of the problem, and these aspects interact with one another. For example, adversary-induced uncertainty might exploit an analyst's inherent cognitive bias, thus leading to an invalid experiment design. Because we are dealing with aspects of a situation, this paper does not refer to our classification as a *taxonomy*, instead using the more general term *ontology* [27][50, pp308,440].

Table I lists these types of difficulty along with interdisciplinary examples, which we will now describe in more detail. This paper is intended to function as a survey and as a reference work so whenever possible, definitions are provided from authoritative sources with page-numbered citations.

### A. Tool

We've already considered that when we perform our malware analysis we might be doing so with buggy programs. In general, no tool is guaranteed to be perfect. Engineers and technicians who create tools to measure physical properties [44, pp29-31] developed measurement theory, which includes several fundamental concepts relevant to cyber security. For example, imagine we are measuring a physical property such as size or voltage. Whatever tool we use will have some **instrumental measurement error** caused by "imperfections" in the tool. So *inaccuracy* is "the imperfection of measurements," and *uncertainty* is this imperfection expressed numerically "with a corresponding confidence probability" [44, p62].

TABLE I
AN ONTOLOGY OF COMPLICATIONS FOR PREDICTION IN COMPUTER SECURITY

| Aspect | | Interdisciplinary Examples: |
|---|---|---|
| **Tool** | | *instrumental error* as in measurement theory [44, p20], *unreliability* as in software reliability [41, p10] and systems engineering [5, p362] |
| **Method** | | *invalidity* [54, p38] as in experiment design, *incorrect specifications* as in operations research [18, pp54-55], *unresolved foundational issues* in statistics [60] [46] |
| **Analyst** | | *personal error* as in measurement theory [44, p20], *errors* as in human factors engineering [63, pp310-315], *biases* as in intelligence analysis [24] [25], *organizational error* as in business management and organizational psychology [17] [20] |
| **Adversary** | | *unpredictability* as in strategic theory [61, §2.2], *induced uncertainty* as in intelligence analysis [24, p.xx], *adversarial invalidation* as in computer security [47] |
| **Context** | • unknown at the moment | *epistemic uncertainty* as in risk analysis [3, pp33-34], *knowledge uncertainty* as in artificial intelligence [49, pp463-464] |
| | • unknowable in practice | *aleatory uncertainty* as in risk analysis [3, pp33-34], *inherent uncertainty* as in intelligence analysis [24, p.xx], *friction* as in warfare [61, §1.7], *experimental uncertainty* [10, p15] as in experiment design |
| | • inherently unknowable | *unknowability* as in operations management [43] [42, p1013] and as in epistemology [45], *deep uncertainty* as in policy analysis [32, pp3-4] |

The term *error* refers to the components of uncertainty, and is usually used with additional adjectives so that we may speak of the difference between "absolutely constant elementary errors," "conditionally constant errors," and "random measurement errors" [44, p63]. Measurement theory identifies and and quantifies these errors using techniques such as calibration measurements against a known value, statistical comparisons of sets of measuring tools, or the use of "indirect data" [44, pp20,57-60].

But beyond the hardware level, computer security uses tools increasingly unlike the physical comparators and transducers common to measurement theory. We must therefore look to the study of *software quality* [62] which focuses on computer programs more like those used in the malware analysis example. One helpful concept is that of *software failure*, which is "the occurrence of an incorrect output ... with respect to the specification" [41, p2]. In contrast, software *reliability* is "the probability that the system will perform its intended function under specified design limits" and **unreliability** is the probability that the software will fail by a given time [41, p10]. An example of software failure in our malware analysis example is if our disassembler produces code that is inconsistent with the language given the input binary; determining the likelihood of this happening is quantifying the disassembler's software unreliability.

It may seem strange to think about the likelihood of software failures – after all, if we find a bug in a program we should fix it and be bug-free. But the challenge occurs with undetected bugs: when testing suggests that our program is bug-free even though the program bugs still has bugs. Even in these cases, software reliability can use statistical methods to produce estimates of the likelihood of undetected errors [35].

Security involves more than software, and security tools are more than software tools. The discipline of systems engineering, which studies complex systems of humans and machines, defines unreliability in the same way as software reliability [5, p362] and indeed appears to be the basis for that definition. More generally, the complexity of systems has led to taxonomies that distinguish between, for example, reliability, dependability, and trustworthiness [57, p1247] and ways of measuring these properties [37].

*B. Method*

Even if our software is bug-free, we still might make the wrong types of conclusions after using it. How do we know if our malware analysis generalizes to other systems? Do we need to vary our testbed, and if so how many variations do we need to use? If we end up with a big set of data, how do we express this data and what types of inferences can we draw from it?

In Section II-C we will discuss errors of reasoning on the part of the analyst. Using an inappropriate method may be an analyst error, but to understand the nature of the error we need to understand the invalidities and uncertainties of the available methods, which are the focus of this section.

We will be guided by the *mechanistic* perspective on experimentation in computing, which defines mechanisms as organized entities and activities that produce regular changes in a process [22, p443]. These mechanisms can be *natural* to the physical world, such as the cognitive aspects of human-computer interaction or *engineered* meaning artificial, such as

DNS schemes and DNS cache poisoning attacks [22, p444-446]. We will consider methodologies for studying these mechanisms, one at a time.

In our malware analysis example, the malware may depend on a human victim clicking on a link sent to them by email, so we may be interested in the *natural mechanisms* of the human tendency to do so. So we might study the likelihood of people clicking on an email link given various types of email text, or after various types of security training.

Methods of studying natural mechanisms include randomized or non-randomized experiments, natural studies comparing events and conditions, correlational observational studies, and meta-analyses [54, pp12,426]. **Invalidity** is when we fail to maintain "the truth of, correctness of, or degree of support for an inference" in the context of a method, where truth is defined in terms of its correspondence to the world, its coherence with related claims, or its pragmatic value [54, pp34-37,513]. There are numerous threats to validity.

After performing a study of humans clicking on potentially-malicious links, if we ask "would a person at another organization also act this way?" we are questioning *external validity*. More formally, external validity refers to "whether the causal relationship holds over variations in persons, settings, treatment variables, and measurement variables." An inference may be invalid if an effect that was found experimentally does not hold when treatments, settings, or observations are varied, for example [54, pp21,82-93,507].

In our example link-clicking experiment if we ask "is this experiment actually capturing what we want?" we are questioning *construct validity*. More formally, construct validity pertains to "making inferences from the sampling particulars of a study to the higher-order constructs they represent." An inference may be invalid if participants react differently in the experimental setting than they would in real life, or act in ways they think the experimenter wants, or if the higher-order constructs are not properly specified, for example [54, pp20,64-82,506].

In our example link-clicking experiment if we ask "were the experimental variations we introduced really the cause of how people behaved?" we are questioning *internal validity*. More formally, internal validity pertains to "whether the relationship between two variables is causal." An inference may be invalid if an effect may be due to systematic participant selection or attrition, or if an effect could have been caused by events not being measured, for example [54, pp53-63,508].

In our example link-clicking experiment if we ask "are the numbers really good enough to suggest a link between a variation and how people behaved?" we are questioning *statistical conclusion validity*. More formally, statistical conclusion validity pertains to " covariation between two variables." An inference may be invalid if a statistical test's assumptions have been violated, if a test has low statistical power in our experiment, or if a test is repeated to "fish" for a statistical significance without correction, for example [54, pp42-52,512].

Informally we may speak of a "valid experiment," but more formally validity relates to inferences rather than methods. Validity depends on circumstances, so there is no methodology that guarantees validity, and determining the validity of an experiment involves "fallible human judgment" [54, pp34-38]. Realistically speaking it is impossible to guard against all potential invalidities simultaneously, so for any given experiment, researchers need to make tradeoffs and be explicit in describing mitigations in their experiment design, and readers need to be aware of potential limitations of an experiment's inferences [54, 96-102].

In our malware analysis example, the malware being analyzed may also depend on a certain type of privilege escalation, so we may be interested in the *engineered mechanisms* which would allow the malware to perform such an escalation. In this case we might study the malware in various testbeds using various instrumentation, for example.

So in this case we are less interested in supporting or falsifying general theoretical principles as with the physical mechanisms above. Rather, we want to know more about *this* particular engineered mechanism given *this* particular engineered environment. This is much closer to the test and evaluation (T&E) approach in operations research. A *test* "involves the physical exercising (i.e., a trial use or examination) of a component, system, concept, or approach for the sole purpose of gathering data and information regarding the item under test." An *evaluation* "is the process of establishing the value or worth of a component, system, concept, or approach" and is generally dependent on the results of a test [18, pp3-4].

One of the first possible errors in conducting a T&E is an **incorrect specification** of the problem [18, pp54-55]. The use and operational environment that is planned for the system under test must be fully understood, including systems that it is expected to work with. If this is not done correctly, then the T&E will provide useless or misleading results. More generally, the first step of operations research is to perform a *system analysis* defining the problem and identifying the appropriate methods to be applied [52, pp301-310][58, pp8-9]. Some of the key components of such a systems analysis include developing descriptive and normative scenarios, preparing objective trees, developing and ranking alternative solutions, and iterating over this process while consulting with stakeholders [52, pp29,57].

Statistical analyses are common to experimentation and T&Es. Any approach that uses data will eventually use statistical and probabilistic techniques, and there are a number of ways of misusing these methods, as suggested above in terms of statistical conclusion validity. But statistics additionally has **unresolved foundational issues**.

In terms of our malware analysis example, imagine we say there is an *n%* probability that the malware will affect our organization's central system. Are we saying that if the malware attacked the central systems 100 times, it would succeed n times? This corresponds to the *frequentist* perspec-tive associated with classical statistics, which holds that a probability is interpreted as "the frequency or propensity of the occurrence of a state of affairs" [46, §2]. Alternately, are we saying that if the malware attacked the system 1 time, our degree of belief in its success is *n%*? This corresponds to the *epistemic* perspective associated with Bayesian statistics, which holds that a probability is interpreted as "the degree of belief in the occurrence of the state of affairs" [46, §2]. But there is more to it: for example, even frequentists may use epistemic probabilities, and epistemic beliefs may relate to an abstract rational agent's beliefs, an abstract decision-theoretic action, or a logical inference under uncertainty.

Related to the frequentist-epistemic distinction is the philosophical *problem of induction*, which considers "the justification of inferences or procedures that extrapolate from data to predictions and general facts" [46] [60]. In our malware analysis example, even if everything remains the same, we cannot *prove* that just because we have seen an outcome once, we will see that outcome again in the future. Perhaps we should only characterize our conclusions as plausible and only seek their *falsification*. But even this is problematic because it requires perfectly-specified causal claims and theory-neutral observations [54, pp15-16].

These are just a few of the disagreements over fundamental questions in the theory of statistics and experimental methodology [54, pp24-29]. But clearly they have not crippled the endeavors of science and engineering. Even philosophers of statistics are less interested in the justification of indiction and more interested in ensuring that statistical methods are properly used [46, §1]. Practically speaking, inherent methodological limitations and foundational uncertainties remind us that our methods give us possibly-useful results rather than guarantees [54, pp30-31].

Every discipline develops an accepted set of methods, so it is natural to conclude this section by asking about the accepted methods of cyber security.

Cyber security researchers have discussed the importance of experimental methodology [34] [48] [56] and the need for foundations [19] [15] [36] but there has been little analysis of these foundations [21], nor very much systematic application of these foundations to research or to applied tasks [6] [23].

However, a discipline's accepted methods change over time [1, §4] so it is possible that the current situation will eventually improve.

*C. Analyst*

Imagine we are performing our malware analysis, and there are no problems with our tools, and our methodology does not lead us astray. There is yet another danger: we ourselves might make a mistake. We might click on the wrong button, or enter the wrong number, or forget to run an important program.

In measurement theory these are called **personal errors**: "individual errors that are characteristic of that person" [44, p20]. Human factors engineering has a number of ways of classifying these errors, but for our purposes we will be guided by an information-processing perspective [63, pp310-315].

In our example malware analysis, if we are given a binary to analyze, and if (unknown to us) it is ransomware, and if after identifying the binary's capabilities we determine that it is a relatively harmless pop-up advertising program, we are making a *mistake*. More formally, mistakes are errors "of interpretation or of formulating intentions." Among other things, a mistake may be due to lack of knowledge, poor communication, working memory overload, exceptions to a common rule, or because of a cognitive bias as described below [63, pp312-313].

In our example malware analysis, if we intend to type "0x00BA" but instead we type "0x00AB" we are making a *slip*. More formally, a slip is when "the understanding of the situation is correct and the correct intention is formulated, but the wrong action is accidentally triggered." Another example is clicking on the wrong button [63, pp313].

In our example malware analysis, if we forget to create a system checkpoint before performing dynamic analysis, we are committing a *lapse*. More formally, a lapse is a failure to perform an action that we intended to. This is commonly called "forgetfulness" with the caveat that human factors engineers specify that a lapse is not due to working memory overload [63, pp313-314].

In our example malware analysis, if we run our binary in the wrong type of introspected environment, we are committing a *mode error*. More formally, mode errors occur "when a particular action that is highly appropriate in one mode of (typically computer) operation is performed in a different, inappropriate mode." Typing a password into a 'user id' field is a simple example [63, pp314].

Personal errors in a system may be identified by a human reliability analysis and may be remedied by careful task design, equipment design, and by training, among others [63, pp315-320].

One approach to understanding and minimizing human error is the intelligence community's study of cognitive **biases**. Biases are "mental errors caused by our simplified information processing strategies," but these are not cultural or political prejudices: "a cognitive bias does not result from any emotional or intellectual predisposition toward a certain judgment, but rather from subconscious mental procedures for processing information" [24, p111].

*Biases in evaluation of evidence* include a tendency to favor "vivid, concrete, and personal" over abstract information, failures in appropriately dealing with missing information, being overly focused on consistency when dealing with limited information, managing possibly inaccurate information, and a persistence in being affected by discredited information [24, pp115-126].

*Biases in perception of cause and effect* result in favoring causal explanations, as well as explanations that include centralized direction, similar causes and effects, and overesti-mating internal rather than external causes, for example [24, pp127-146].

*Biases in estimating probabilities* include being overly influenced by available examples or an initial approximation, or misinterpreting subjective or quantitative expressions of probability [24, pp147-160].

*Hindsight biases* include overstating the accuracy of past judgments, underestimating how much is learned by those who read reports, and an after-the-fact tendency to believe that events were more easily forseeable [24, p161].

These biases are the result of human cognition and as such are not always easy to be rid of. However, they can be managed by structured analytic techniques [26] such as the Analysis of Competing Hypotheses [24, pp95-110].

So far we have only considered errors that are caused by an individual analyst, but errors can also be caused by sets of analysts in an organization. In the disciplines of business management and psychology these **organizational errors** are an active area of research, so there are many approaches possible [17] [20]. We consider two examples that seem particularly applicable to our domain.

Our first example comes from a defense analysis study of *organizational failure*, specifically defined as relating to an avoidable defeat: "failures attributable neither to gross disproportion in odds nor egregious incompetence on part of the victim nor yet to extraordinary skill on the part of the victor" [9, p.*v*]. From this perspective there are three basic failures that can either exist on their own or combine together [9, pp25-26]. A *failure to learn* is when an organization is aware of a danger but does not appropriately observe and react to that danger, leading to the organization suffering from the results of that danger themselves. A *failure to anticipate* is when a danger has not been observed but is easily predictable, and yet the organization does not appropriately react. A *failure to adapt* is when circumstances are rapidly changing but which could have been minimized by reacting appropriately and planning beforehand. These could be better understood and avoided by analyzing historical case studies, exploring the various tasks involved by command level, and determining the types of failure that occurred and their effects.

Our second example[1] comes from an information warfare perspective: *organizational malfunctions* are when an attack causes a decision-making operation to "operate in a way that was unintended by the original designer of the organization and is probably undesireable for the organization." The following characterization is developed from technical analogies, uses examples from the military, political, and business decision-making, and focuses on "malfunctions that arise specifically because of the large, distributed nature of decision-making organizations and attendant systems."

From this perspective, a *Tardy Decision* is made when required information arrives too late to be used; a *Low and High Threshold* relates to the amount of evidence required

---

[1]The categories and quotes in this example are from Kott [29, pp115-134]

---

for a decision: a low threshold of information required for decisions can result in too many decisions being made; a high threshold can result in not enough. An *Excess of Timidity or Aggressiveness* occurs when a reaction is too weak or too strong given the circumstances. A *Self-Reinforcing Error* occurs when an action is made in the wrong direction, serving to increase the situation's problems. An *Overload* is when a component of the system has met a limit and further increase has no effect. A *Cascading Collapse* is when an inappropriate decision in one part of the system propagates to other parts of the system. A *Misallocation of Authority* is when decision-making is decentralized in suboptimal ways. A *Lack of Synchronization and Coordination* is when there is a misalignment in decision-making. A *Deadlock* is when (for example) a system component is waiting for a resource from another component, which itself is waiting for a resource from that first component. Finally, *Thrashing and Livelock* is when a system is unable to reach a decision because it spends too much time on action that is ultimately unproductive, possibly because those actions look productive at the local level.

*D. Adversary*

The malware sample in our malware analysis example has several ways of fighting back against us. The malware binary might be encrypted or obfuscated, it may attempt to deceive us as to what it is really doing, or it may change its behavior if it believes it is being observed as it runs [55]. If the malware authors come to believe that their malware has been detected, they may change how the next iterations of their malware works. Or they might change how their malware works just in case anyone has detected them, without even being sure if that is the case. So any predictions we make about the malware's behavior is contingent on its antagonistic nature.

In strategic theory this **unpredictability** is due to the fact that an action "must expect positive reaction" and "the very nature of interaction is bound to make it unpredictable" [61, §2.2]. It's worth looking more closely at this characterization, which is from Clausewitz. Any action we make during a conflict "must expect" a reaction – this is by definition, because we are involved in a conflict with an intelligent reactive adversary. That adversary may react through a "negative action" to avoid or mitigate the results of our action, or a "positive action" to advance their cause at our expense.

"The very nature of interaction" is that the adversary also expects reactions from us, so are motivated to make unpredictable actions. The best outcome for the adversary would be to achieve *surprise*, in which we mis-estimate our adversary but do not realize this until after we are attacked [12, p18-note59]. To achieve this, the adversary may conduct denial and deception operations to create a "fog" of **induced uncertainty** [24, p.xx].

In strategic theory and intelligence analysis, *deception theory* describes the ways that an adversarial organization may attempt to induce uncertainty for their benefit. There are many types of theories, but they tend to have certain principles in common: a deception operation works in the context of the truth; this truth is denied to the target; deceit is used; the target is misdirected [4, pp58-66]. These principles can be mapped in a 2x2 matrix as shown in Table II. The "concealing a fiction" and "revealing facts" quadrants may seem counterintuitive, but are important in building credibility during a deception operation. In fact, deception operations should be carefully managed, with specific objectives defined, information presentation and adversary behavior tracked [4, pp49-57]. And to defend against adversary deception operations, there are a variety of technical and operational counter-deception techniques [4, pp143-305].

TABLE II
PRINCIPLES OF DECEPTION AS DESCRIBED BY BENNETT AND WALZ AND APPLIED TO CYBER SECURITY [4, pp51-54,124]

| | |
|---|---|
| **revealing** a *fiction* as with MITMs and honeypots | **concealing** a *fiction* e.g. OPSEC to withhold operational capabilities |
| **revealing** *facts* as in releasing limited capabilities to reduce attacker sensitivity | **concealing** *facts* as with cryptography and stenography |

These ideas are relevant to our malware analysis example because security is fundamentally *operational* meaning it depends on a specific context involving missions and adversaries; so the malware we are analyzing may be part of a larger deception operation, and understanding this may be important for a proper analysis [47]. Security's adversarial and operational characteristics affect the types of methods that we may use. The experiments and tests described in Section II-B are fundamentally different from methods that involve tactical methods used against an adversary, but these methods against an adversary are sometimes informally referred to as "experiments" or "tests."

An example of this is the historical "AF ruse" that occurred during World War 2 just before the Battle of Midway [12, pp57-59]. Naval intelligence had decrypted their adversary's messages referring to a target named AF, but there was some doubt about whether AF referred to the Midway military installation. To test this, naval intelligence sent a message to Midway via undersea cable (which could not be intercepted) ordering them to report back, via unencrypted transmission, about a problem with their water supply. The unencrypted transmission was duly sent, and several days afterwards an adversary message was decrypted to reveal their belief that AF had a problem with their water supply. This ruse confirmed that AF was the code name for Midway.

Although the AF ruse involved technology and natural phenomena, it was neither an experiment nor a test as defined in Section II-B. Formally, the AF ruse was a tactical deception operation to determine the state of an adversary's knowledge. The underlying motivation for the AF ruse was a disagreement within naval intelligence about the identity of AF [12, pp57-58]. The AF ruse was therefore also an operation to reduce the type of analyst organizational uncertainty described in Section II-C.

Of course, there was always the possibility that the AF ruse had failed: that the adversary knew their messages were being read and that the adversaries themselves were being deceptive, perhaps because they had been informed via espionage. This would be an example of induced uncertainty through deception. Similarly, the adversary may have developed an undetected way to read the undersea cable; this type of uncertainty, related to the development of technology, is what we shall consider next.

From the perspective of military history we can consider examples of technology's effect on conflict. Analyses of past events tend to focus on either the development of new technology, or the effect on conflict once a new technology is introduced; however, a "problem-solution-application" approach integrating both of these is more complete [28, pp39-40]. When a belligerent has a problem related to their conflict, they develop a (possibly technological) solution and apply it to the conflict. But this creates a problem for their adversary, who then seeks a solution (also possibly technological) and applies it. Which creates a new problem for the original belligerent, and so the adversarial cycle continues.

There is "chance and fortune" in this process [28, p50], though organizational quality (as described in Section II-C) plays a role. More relevant to the Adversarial aspect of uncertainty is the following. Imagine a belligerent sees that their adversary is using certain technology successfully against them. The belligerent manages as best they can, while pursuing technology to enable more effective counter-tactics. The belligerent's technology arrives, but in the meantime their adversary has also been pursuing their own next generation of technology. The extent to which the original belligerent's new technology has been made irrelevant by their adversary's new technology is **adversarial invalidation**: when a "method has been undermined or counteracted by the antagonist and is no longer as useful" [47]. Both the first belligerent and their adversary will try to outguess each other both in developing technology, as well as in the planning of tactics and operations[2] that use this technology. This is one of the motivations for the importance of understanding your adversary, which has been emphasized since at least the time of Sun-tzu.

There are several technical approaches to representing and reacting to adversary actions.

---

[2]Where *tactics* are the use of forces in direct contact (e.g. on a battlefield), and *operations* are related sets of tactical actions [2, p13] [13].

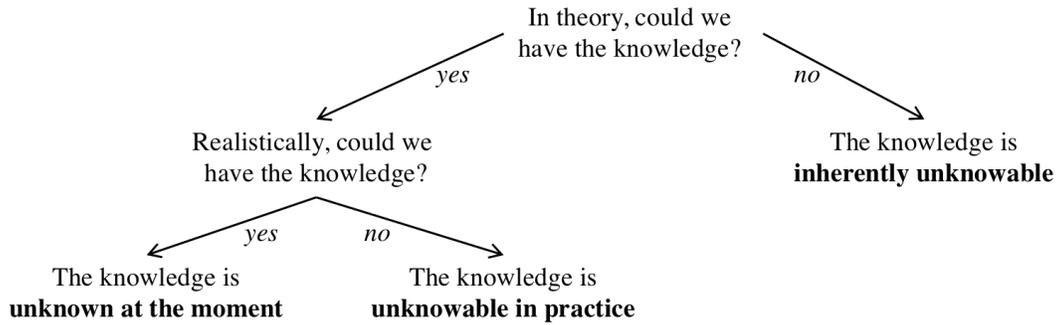

Fig. 1. Characterizing Contextual Unknowns

*Game theory* allows systematic analyses of "interactive decision making" [14, p5]. These interactions are represented as abstract strategic games, generally between two participants, where the games can be sequential vs simultaneous, in total conflict vs some commonality, once or repeatedly, with the same or with changing opponents, with full information or not, and with equal information or not, and with fixed, manipulable, or enforceable rules [14, pp18-27]. Analysis of these games can provide explanations of events, prediction about what types of outcomes are possible, and prescription of the types of strategies likely to lead to desireable outcomes [14, pp36-37].

In artificial intelligence, competitive environments such as two-player games can be analyzed through *adversarial search*: given a starting situation, creating a search tree for all possible moves. This search tree is often too large to exhaustively search, so pruning and search heuristics are usually applied [49, pp 161-185]. AI methods are often combined together. For example, neural networks can be used to evaluate moves, with their weights trained by evolutionary methods or by reinforcement learning [33, p69]. A variety of AI techniques have also been used to model the adversary: identifying their objectives, their most likely future behaviors, and their current plan [31, p77].

Operationally, the military uses *analytic wargaming*[3] for "the exploration of the role and potential effects of human decisions" [40, p164]. Rather than attempting to make a prediction or a generalization, a wargame should have a particular objective, and it should generate data that allows the study of how the players interacted and made decisions [40, pp270-271]. Such wargames are also good for examining multiple possible outcomes of conflicts [51, pp55-56].

### E. Context

In our malware analysis example, even if our tools have no errors, if our methods are not problematic, if we make no personal errors and if our adversaries do not outwit us, there still may be some unknowns in the environment. The malware may exploit features in the host system that we were unaware of, for example.

Our ontology will be guided by concepts from risk analysis and operations management to distinguish between three types of contextual unknowns as shown in Figure 1: those that are knowable in theory but unknown at the moment, those that are knowable in theory but unknowable in practice, and those that are inherently unknowable.

*1) Unknown at the Moment:* When we begin the malware analysis of our motivating example, the malware has a number of properties that we have not yet discovered though we eventually will. The tests and experiments described in Section II-B are conducted to reduce these unknowns.

In risk analysis, "those things about which we could learn if we were able" produce **epistemic uncertainty**, also described as "subjective, type B, reducible, state of knowledge" [3, pp33-34].

Similarly, measurement theory refers to errors that "arise as a result of an inadequate theory of the phenomena on which the measurement is based and inaccuracy of the relations that are employed" [44, p20]. This is referred to as "methodological error" though it corresponds to the aspect of *context* in our ontology, rather than methods as described in Section II-B.

Another similarly is in **knowledge uncertainty** in artificial intelligence [49, pp463-464] which has several causes. First *theoretical ignorance* because of "no complete theory of the domain". Second, *practical ignorance* because, for example, "not all the necessary tests have been or can be run" (if this does not preclude that this theory can be developed or that these tests can be run. If the test cannot be run, it is material for the next section.) Third, *laziness* because "it is too much work" to acquire the relevant knowledge or "too hard to use" that knowledge. The provocative term 'laziness' from the source [49] is presumably not meant to indicate a moral failing; when it refers to situations in which acquiring and using knowledge may simply be computationally infeasible due to run-times, then it is also material for the next section.

*2) Unknowable in Practice:* There may be other malware properties that we could, in principle, discover but which we are unable to, perhaps because we lack the computing power or information about the malware authors. For example, imagine we wanted to know the malware author's immediate surroundings, thinking that this would give us a clue to a password that we need to know. In theory this is knowable, but in practice it is usually not.

In risk analysis, things "about which we either cannot or choose not to learn" produces **aleatory uncertainty**, also described as "stochastic, type A, irreducible, variability" [3, pp33-34].

---
[3]In contrast to *training wargaming* [40, p167].

In engineering, there is an uncertainty that is "variability unrelated to the errors inherent in the measuring systems" and includes, for example, variable flow rates in timewise experiments, sample variability in sample-to-sample experiments, and measurement variability in transient experiments [10, pp15-17]. This is referred to as "experimental uncertainty", though it corresponds to the aspect of *context* in our ontology, rather than methods as described in Section II-B.

It is similar to **inherent uncertainty** in intelligence analysis, "the natural fog surrounding complex, indeterminate intelligence issues" [24, p.xx]. It is also similar to **friction** as in warfare; as Clausewitz says, "Countless minor incidents – the kind you can never really foresee – combine to lower the general level of performance, so that one always falls far short of the intended goal." [61, §1.7].

*3) Inherently Unknowable:* Finally there may be knowledge that we are completely unaware of and which we could not be expected to know about. To talk about such *inherently unknowable* knowledge might seem to imply either "I know the fact *x* but I know it incorrectly" or "I do not even know that the fact *x* exists," both of which are self-contradictions if *x* is the content of the fact. However, a more tenable interpretation is to think about *x* as part of a class of knowledge of which we are unaware, without making claims about its content. [45, p316]

This is called **unknowability** as in operations management [42, p1013] [43]. When managing a project, there are "known unknowns" which include the items that are currently unknown or unknowable in practice, but in either case we are aware that we lack knowledge about it. However, project managers should also expect "unknown unknowns": the material that we do not know, and whose possibility we are only dimly aware of. This is related to *Black Swan events* as described in popular literature, which refers to unexpected events that are outliers, that have an extreme impact, and that are retrospectively predictable [59, p.xxii]

From the perspective of intelligence analysis, *convergent* phenomena refers to "behavior of any system [that] could be predicted from the average behavior of its component parts" and *divergent* phenomena "are not governed by the laws of cause and effect" or are otherwise unpredictable [8, pp186-187]. This category relates to divergent material.

The discipline of policy analysis is familiar with the unknowability of the future, as it is their task to recommend decisions that will have impacts lasting for decades. In this situation of **deep uncertainty**, there is no clarity on the models to use, the probability distributions of variables and parameters, and the way of valuing potential outcomes [32, pp3-4]. This is challenging because "the future is certain to follow paths and offer events we did not imagine" [32, p8].

Unknowable does not necessarily mean beyond the realm of human comprehension. If an operation suddenly encounters a previously unknown challenge, it is clearly no longer unknown. At that point it likely becomes an unknown at the moment or an unknown in practice, although in some cases the effect may simply be considered contextual *noise* and handled with one of the methods described below.

By definition the entire category of inherent unknowables is ill-defined. The study of the exact nature of our ignorance is sometimes called *agnotology* in the humanities [11]. In epistemology, there is an acknowledgment that "the realm of ignorance is every bit as vast, complex, and many-faceted as that of knowledge itself." [45, p316].

There are a number of technical methods for managing contextual unknowns. In artificial intelligence, *Bayesian networks* uses linked random variables and conditional probabilities to specify the uncertainties about the context [49, pp492-495]. These can then be combined with decision and utility nodes to create *decision networks* [49, pp597-600] to represent the influences on simple decisions. More complex sequences of decisions are represented with a *partially-observable Markov decision process* (POMDP), which tracks probabilistic beliefs about the context through sequences of actions [49, pp613-631].

Analytic approaches include scenario planning methods [7], an example of which is *long-term policy analysis* (LPTA). LPTA uses computer-assisted models to consider hundreds to millions of possible future scenarios. These scenarios are interactively explored by human analysts who seek strategies that will enable robustness across the greatest range of possible futures [32, p.xiii].

Operationally, organizations use open control loops representing tactical/operational-level decision-making, one of the most well-known of which is Boyd's OODA loops [38] [39]. This decision-making uses the following approach: Observe, Orient, Decide, then Act, in this way remaining reactive to any previously-unknown contextual situation or adversarial action. In general, organizations seek to be flexible against contextual unknowns while being resilient against adversarial surprise and reliable against the types of errors described in Section II-C [16].

III. CONCLUSION

This paper makes several contributions relevant to the foundational problem of prediction in cyber security.

First, it approaches the problem from the angle of complications to prediction. This allows a systematic interdisciplinary study of the types of complications involved, including error, unreliability, unpredictability, and more. The same term can be used differently in different disciplines, so definitions are provided from authoritative sources, with page-numbered citations for easy reference and further consideration.

Second, an ontology is developed to describe the various aspects of the problem. This allows a better understanding of the aspects of the challenge, in non-jargon language that is nevertheless tied to the referenced disciplines. These aspects are explained in terms of a malware analysis example scenario, to describe their interdisciplinary methods to the cyber security context.

Finally, along with the complications to prediction, we have also described the methods that the various disciplines have used to mitigate, express, or react to these complications. These are described as "Consolations" and described in a very general way, because the contextual and adversarial nature of cyber security guarantee that these ways of managing complications cannot universally solve the challenges to prediction; they can only improve the situation, and perhaps make it winnable.

As a final thought on the malware analysis example: much can go wrong when making a prediction. Perhaps our ultimate consolation is that we are not just building structures in, or developing knowledge about, an indifferent universe; we are actually trying to outwit adversaries who are themselves also making predictions. And these adversaries, though craftier than an unthinking Nature, are as limited as we are, by the very same complications.

ACKNOWLEDGMENTS

Thanks to Jeremy Blackthorne for relevant conversations, and to John Wilkinson for his support and encouragement.

This material is based upon work supported under Air Force Contract No. FA8721-05-C-0002 and/or FA8702-15-D-0001. Any opinions, findings, conclusions or recommendations expressed in this material are those of the author(s) and do not necessarily reflect the views of the U.S. Air Force.

Distribution Statement A. Approved for public release: distribution unlimited.